# The CeMIn$_5$ (M = Rh, Ir and Co) System and the Single Energy Scale Model of Metamagnetism

A.Thamizhavel[1], B.D. White[2], M.B. Maple[2], S.Ramakrishnan[1], Ulrich Welp[3], V.Celli[5], Pradeep Kumar[4] and Bellave Shivaram[5]

[1]Tata Institute of Fundamental Research, Homi Bhabha Road, Mumbai, India.

[2]Department of Physics and Center for Advanced Nanoscience, University of California, La Jolla, San Diego, CA. 92093.

[3]Materials Science and Technology Division, Argonne National Labs, Lemont, IL. 60637.

[4]Department of Physics, University of Florida, Gainesville, FL. 32611.

[5]Department of Physics, University of Virginia, Charlottesville, VA. 22901.

## ABSTRACT

For a decade and a half, CeCoIn$_5$ and related alloys have served as a rich playground to explore the interplay between magnetism and unconventional superconductivity. Despite this extended study, the presence/absence of metamagnetism (MM) in this ternary system remains as an unresolved issue. Here we show that the linear and non-linear magnetic response in CeMIn$_5$ (M = Rh, Ir and Co) can be understood within the context of the recently proposed single energy scale (SES) model of MM. New measurements of the third-order susceptibility, $\chi_3$, in CeCoIn$_5$ are presented and together with the known systematics of the linear susceptibility in all three compounds, are shown to be consistent with the SES model. Predictions are made for the MM critical field in CeCoIn$_5$ and the fifth-order susceptibility, $\chi_5$.

Strongly correlated electronic systems in general and the heavy electron family of metals in particular exhibit a rich interplay of magnetic and superconducting responses that are easily tuned with pressure or composition[1]. In the CeMIn$_5$ family of alloys, it is possible to access a variety of different ground states through tiny changes in the lattice parameters and their anisotropies[2]. In CeCoIn$_5$, robust superconductivity is observed at $T_c$=2.3 K under ambient pressure[3], whereas in CeRhIn$_5$ an antiferromagnetic ground state is observed which gives way to superconductivity at a low pressure of 0.6 GPa. Superconductivity also appears for M = Ir, with a relatively low $T_c$=0.4 K, but with no evidence of long-range magnetic order (or any other type of broken symmetry)[4]. A large part of the interest in this ternary system has been driven by a desire to decipher the precise nature of the connection between the easily tunable and coexisting superconducting and magnetic phases[5]. In this work, we confine ourselves to metamagnetism (MM) or rather the absence/weak nature of MM in the CeMIn$_5$ system. The question of the relationship of MM to unconventional superconductivity is also a relevant one, but is outside the scope of the present investigation.

Metamagnetism is a phenomenon commonly observed in heavy fermion compounds and other strongly correlated systems. Here a magnetic field causes a rapid rise in the magnetization at a critical field where the rise in magnetization becomes sharper as the temperature is reduced. Concomitantly a peak is observed in the linear susceptibility at a specific temperature in many materials. A linear correlation between the temperature of the peak, $T_1$, and the critical field, $H_c$, has also been established[6]. This scaling along with a similar correlation between the temperature where a peak in the third-order susceptibility (also seen in many heavy fermion materials) occurs[7] has lead to a single energy scale (SES) model of MM[8]. This model has also been augmented recently to account for the large non-zero value of the linear susceptibility at T=0, a feature common to heavy fermion materials as well as the large Curie-Weiss constants that are simultaneously observed[9]. Very often, MM is extremely anisotropic – it is observed with the magnetic field with respect to the crystalline axes in specific directions only. In the direction where there is no metamagnetism the material behaves almost like a paramagnet. The SES model, even though it is a "single-site" model, captures all of these key features of correlated metamagnets.

The magnetic properties of CeMIn$_5$ at first glance appear to violate the standard model of a correlated metamagnet painted above. As stated earlier, MM in the CeMIn$_5$ system is almost non-existent. In CeCoIn$_5$, there is no MM transition; nevertheless, there is a sharp rise in the magnetization very close to the upper critical field $H_{c2}$, and this feature is usually attributed to superconductivity[10]. At higher fields (23 T), an anomaly in the Nernst effect has been identified whose 'thumbprint' is very similar to that seen in CeRu$_2$Si$_2$, a well known metamagnet[11]. However, there are no associated anomalies in the magnetization in the same field range. In CeIrIn$_5$, there is weak MM at 42 T when the field is along the hexagonal c-axis and a possible MM transition at 50 T in the perpendicular case[12]. In contrast, in CeRhIn$_5$ there is no MM along the c-axis for fields up to 55 T, but there is a weak one in the perpendicular direction[13] at 2 T. A a peak in the linear susceptibility, another characteristic signature of a metamagnet, isabsent in both CeCoIn$_5$ and CeIrIn$_5$.  While χ$_1$ does have a peak in CeRhIn$_5$, with T$_1$=6 K,   it is marginal and the zero-temperature susceptibility is almost equal to its value at the peak. Thus as a good approximation, we can state that MM as observed in other heavy fermion compounds such as UPt$_3$ and CeRu$_2$Si$_2$ does not exist in the CeMIn$_5$ family of materials.  It would, therefore, appear that any attempt to understand the properties of this system within the standard framework we have developed to understand MM in heavy fermion materials and other strongly correlated systems would be a futile exercise.

However we will demonstrate below, using new results for the third-order magnetic susceptibility in CeCoIn$_5$, taken together with the earlier established behavior of the linear susceptibility, that magnetism in all three members of the CeMIn$_5$ family can be well understood  within  the framework of the SES model of MM. Through our analysis, we also provide predictions for the behavior of χ$_3$ for M = Ir and Rh and the fifth order susceptibility, χ$_5$, for all three cases.  These predictions can be checked in a straighforward manner through further experiments.

For the work presented here, we used two batches of single crystals of CeCoIn$_5$ - one synthesized at the University of California, San Diego and a second one at the Tata Institute of Fundamental Research, India, both using a flux growth method.  As-grown crystals without annealing after synthesis were used in the measurements.  Magnetization measurements were performed in a SQUID VSM (to 7 Tesla) at TIFR (primarily on sample #2) with a second set of measurements performed on a DC SQUID MPMS (also 7 Tesla) at Argonne National Labs (on sample #1).

In figure 1, we show the linear magnetic response for sample #1 for magnetic field of 1000 Oe parallel applied to the c-axis.  We reproduce the results seen in the very first measurement on this system by Petrovic et al.  There is a Curie-Weiss response at high temperatures which gives way to a region where χ$_1$ is nearly temperature independent (plateau region) in the range 40 K to 20 K, below which, a pure paramagnetic-type response develops. To enable a comparitive discussion that follows, we also show in figure 1 the results on CeRhIn$_5$ from ref. 12.   The weak maximum in this system at ~ 6 K referred to earlier, is apparent in the figure[14].

We model the behavior of CeCoIn$_5$, shown in fig.1, starting from an ansatz.   An examination of the data in fig. 1 suggests that the linear susceptiblity can be obtained approximately by a superposition of the response of a metamagnetic S = 1 pseudospin, which will have a maximum in the linear susceptibility in the neighborhood of 50 K,  and the response of a second spin which behaves as a paramagnet at all temperatures.  This ansatz is illustrated by the dotted lines in fig. 1.  Such a superposed behavior can be modelled effectively with a Hamiltonian that involves <u>two</u> energy scales, Δ$_1$ and Δ$_2$:

$$H = \Delta_1 S_{1z}^2 - g_1 h\, S_{1z} + \Delta_2 S_{2z}^2 - g_2 h\, S_{2z} \qquad (1)$$

Here $g_1$ and $g_2$ are the efefctive g-factors of spins 1 and 2 and h is the magnetic field.

We start in (1) with $\Delta_2$, "the single energy scale", taken to correspond to the MM spin and $\Delta_1$ to correspond to the paramagnetic part. Since this latter part has to dominate the very low temperature behavior, we can guess that $\Delta_1 \sim 0$.

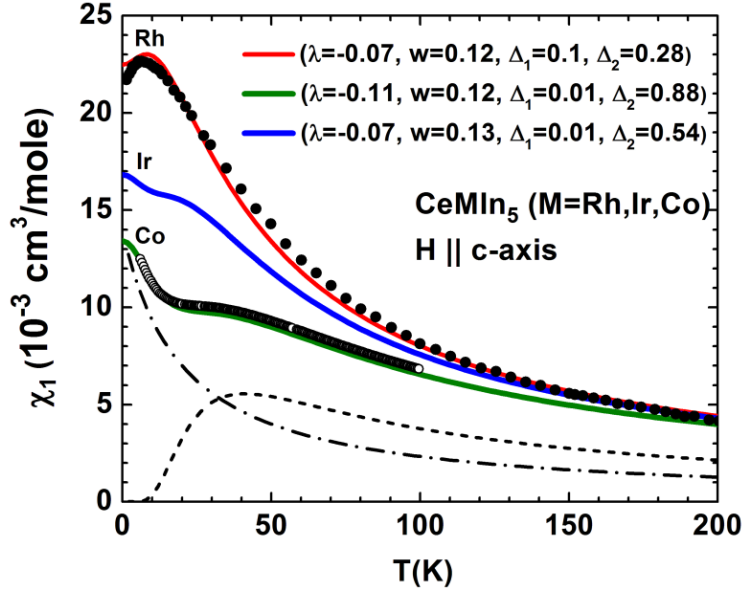

*Figure 1: The linear susceptibility for magnetic field applied parallel to the hexagonal c-axis of crystalline CeCoIn$_5$ (open circles). The solid green line is from the model using the parameters given in the figure. The plateau in the susceptibility between 20 K and 40 K and the rapid rise below 20 K are reproduced very well in the model. The dash and dot-dash lines represent the ansatz of decomposing the CeCoIn$_5$ response to a MM and a paramagnetic part. Also shown are the model results for CeIrIn$_5$ (blue line) and CeRhIn$_5$ (red line). The solid circles are data reproduced from Petrovic et al. (ref.3).*

The Hamiltonian (1) is a simpler version of a more general model with three spins developed by one of the authors (ref. 10). In this general approach, the energy levels, $E_i$, of the spin system evaluated with Hamiltonian (1) are used to compute the thermodynamic quantities via the partition function $Z = \sum_i e^{-\frac{E_i}{k_B t}}$, except that a non-zero width, $w$, is introduced for the energy levels. The effect of this "hybridization" broadening of the levels is to replace the thermodynamic temperature $t$ by $t_w = \sqrt{t^2 + w^2}$. In addition, a mean field parametrized by $\lambda$, which shifts the perceived magnetic field from $h$ to $h+m\lambda$ is also introduced. The values of these parameters, $\lambda$ and $w$, along with the energy levels $\Delta_1$ and $\Delta_2$ for CeCoIn$_5$, are shown in fig. 1. As anticipated the best fit value of $\Delta_1$ for CeCoIn$_5$ is $\sim 0$. In addition, to obtain proper fits[15], we scale the model temperature '$t$' to the experimental temperature $T$ as $T = 85$. We also find in the model that $T_1=(2/3) \Delta_2$ and hence an implied peak in the linear susceptibility at 57 K, a reasonable value, given the behavior of $\chi_1$ as seen in fig. 1. It is obvious that with these model parameters equation (1) describes extremely well the behavior of CeCoIn$_5$. With only minor changes in the parameters equation (1) also describes the dramatically altered magnetic response in the two other materials CeIrIn$_5$ and CeRhIn$_5$, as seen in the figure. This is in line with the experimental observations that small changes in the lattice parameter (without significant changes in the electronic structure of the constituents) enable tuning between very different ground states easily. The single energy scale, $\Delta_2$, however, does increase substantially in going from Rh to Co.

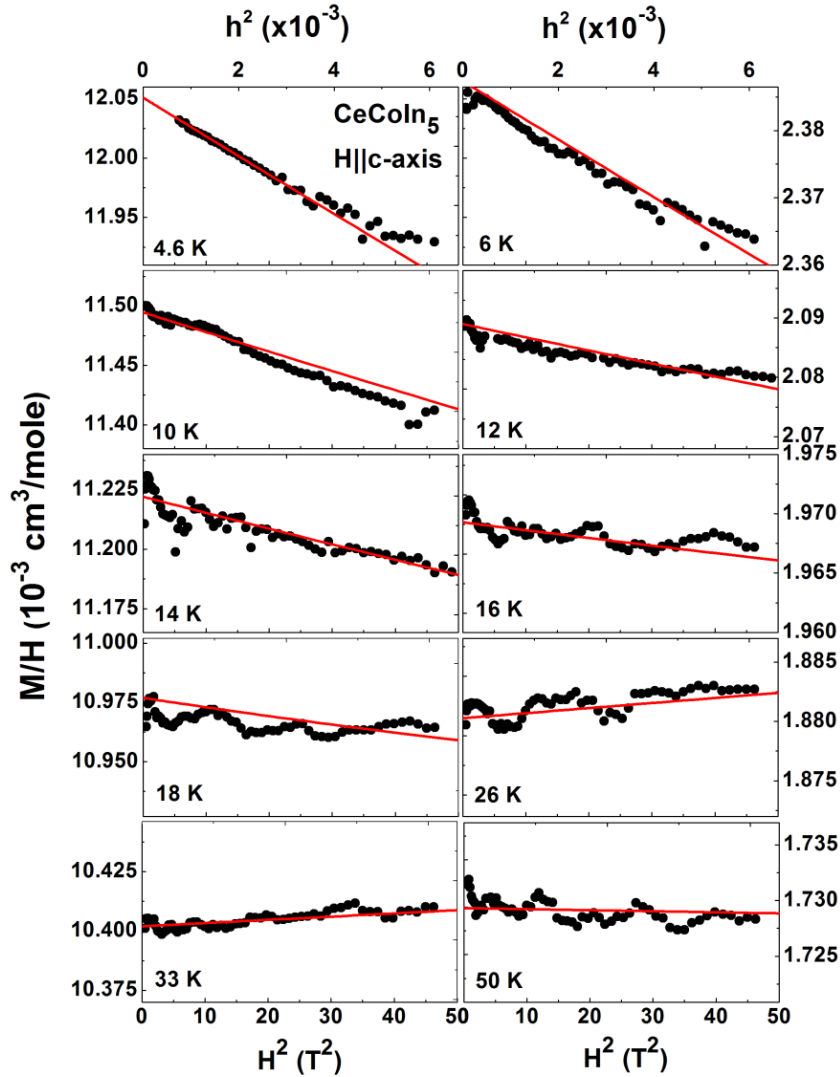

*Figure 2: Illustration of the procedure employed to extract the third-order susceptibility from the measured magnetization isotherms. The slope of the best-fit lines in each of the panels is the third-order susceptibility, $\chi_3$ and the intercept is $\chi_1$. $\chi_3$ is negative at the lowest temperatures and decreases in magnitude as the temperature is increased with a weak positive maximum at ~30 K.*

We next turn to a discussion of the nonlinear susceptibility. Experimentally, the leading order non-linear response is determined by plotting $M/H$ vs $H^2$ and by taking the slope of the resultant straight line as suggested by eqn.(2).

$$M(T) = \chi_1(T)H + \chi_3(T)H^3 + \cdots \qquad (2)$$

Such plots for the CeCoIn$_5$ system when the field is applied parallel to the c-axis are shown in fig. 2 with the extracted values of $\chi_3$ shown in fig. 3. The scaling between the experimental magnetic field $H$ (in tesla) and the model magnetic field, $h$, is given by the same factor as the one employed for scaling the temperatures. This is to be expected since there is a 1:1 correspondence between $T_1$ on the absolute temperature scale and the critical field $H_c$ expressed in tesla in heavy fermion materials empirically. Thus, since the magnetization isotherms were measured to a maximum field of 7 T, the maximum value of $h^2$ in fig. 2 is $h^2=(7/85)^2$. This conversion between $h$ and $H$ also enables us to quantitatively compare the model $\chi_3$ with the experimental slopes from the panels in fig. 2. A seen in fig. 3 this agreement is excellent – the third-order susceptibility is mostly negative except for a weak maximum around 30 K where it has a small positive value. The position of this maximum is also consistent with the well established $T_3=0.5T_1$ scaling(ref. 4) that arises in the SES model. As a further rigorous test of our model, measurements of $\chi_3$ in the other two systems, CeRhIn$_5$ and CeIrIn$_5$, should be carried out. The expected behavior of $\chi_3$ in these two materials is also shown in fig. 3.

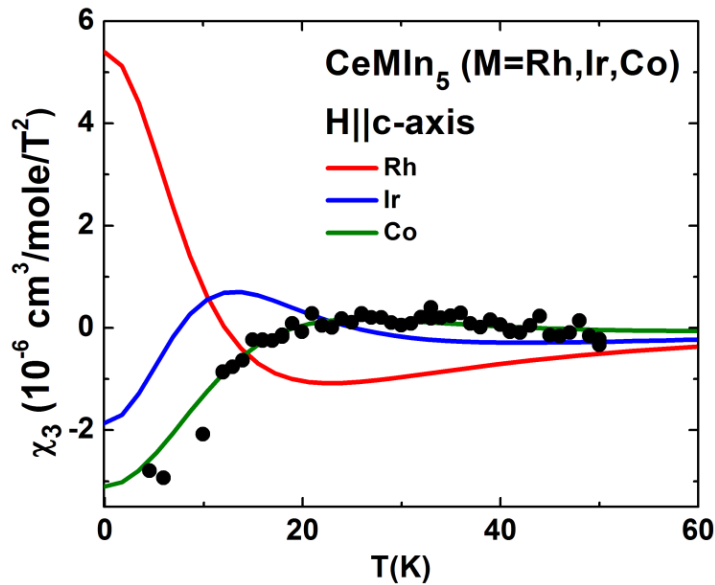

*Figure 3: The measured values of the third- order susceptibility in CeCoIn$_5$ for magnetic field applied parallel to the c-axis (closed circles). The green line shows the model calculation for $\chi_3$ using the parameter values for CeCoIn$_5$ stated in fig. 1. The expected $\chi_3$ values for the two systems, which have not yet been measured, CeRhIn$_5$ and CeIrIn$_5$ are shown in red and blue, respectively. It is remarkable that small changes in the parameters can cause dramatic shifts, from deeply negative to deeply positive, in $\chi_3$ at the low- temperature end.*

Continuing with our approach, we show in fig. 4 the evaluated differential susceptibility for *h||c-axis* extended to higher fields for the three compounds. There are no sharp peaks in either of the three compounds and the susceptibilities start with high values in zero field. Thus, no MM would be expected (or will be very weak) in conformance with experimental observations. A seen in fig. 4 (a weakened) MM in CeIrIn$_5$ occurs at *h=0.45* in good agreement with the experimental value of $H_c$ = 42 T given the scaling between *h* and *H* we established above. No MM would be expected in CeRhIn$_5$ (red curve), also consistent with known experimental observations. However, we do expect to see a weak transition around 75 T- 80 T in CeCoIn$_5$. Verifying this latter prediction as well as testing for the behavior of χ$_5$, the fifth-order susceptibility presented in fig. 5 would be further important tests of our model.

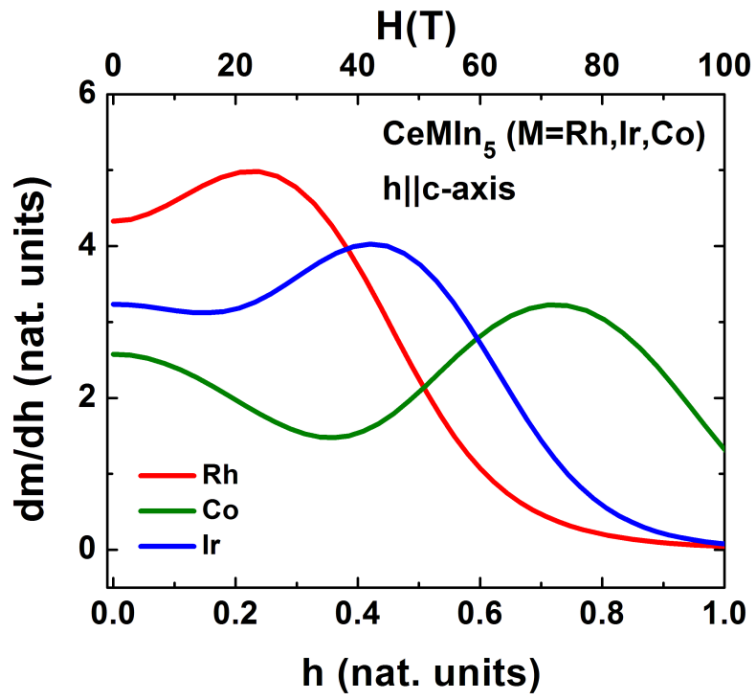

*Figure 4: The calcuated differential susceptibility at high fields in the CeMIn$_5$ system using the model parameters obtained from an analysis of the linear susceptibility - a low field property. The obtained behavior of dm/dh is in good agreement with the known high field measurements. In CeIrIn$_5$ a weak metamagnetic transition is observed at 42 T and this corresponds to the peak at h=0.45 (blue curve). The peak in the red curve is not very pronounced and this is in accord with absence of MM in CeRhIn$_5$. A broad but measureable transition is expected in CeCoIn$_5$ in the 80T range (green curve).*

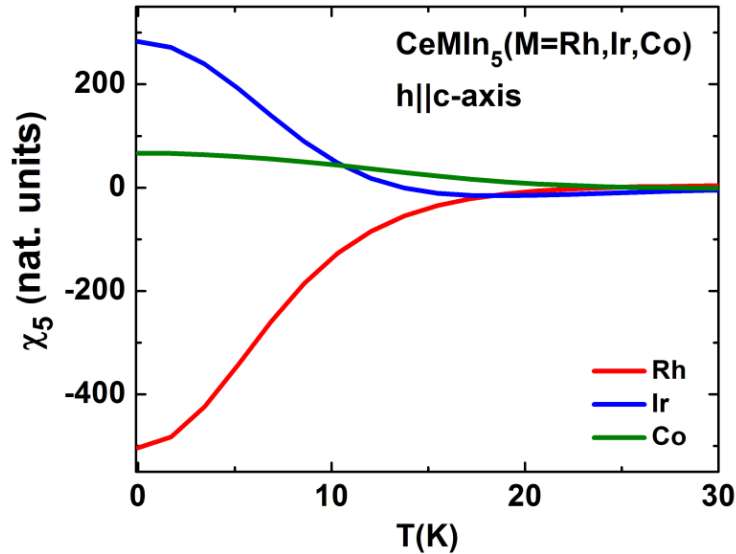

*Figure 5: The model calculations of the fifth- order susceptibility, $\chi_5$, for all three members of the CeMIn$_5$ system. Note the reversal of the expected behaviors in $\chi_5$ compared to $\chi_3$ – for CeRhIn$_5$, it is negative at low T, but for CeIrIn$_5$ it is positive. This sign relationship is reversed for $\chi_3$.*

We next turn to a discussion of the linear susceptibility for the perpendicular orientation, $H||a$-axis. Previous work on the CeMIn$_5$ system indicates that $\chi_1$ is fairly featureless compared to the parallel case and the behavior is very similar to that of a regular paramagnet. However, an approach to saturation as the field is increased is seen only in CeCoIn$_5$. In both CeRhIn$_5$ and CeIrIn$_5$, a strikingly perfect linear response is obtained in fields up to 55 T[16]. It should also be noted that the perpendicular direction is the hard axis with the susceptibility being 1/2 -1/3 of the value in the parallel direction in all three compounds. In our model, the 'bare' susceptibilities (i.e., without the addition of a mean field) are reversed in the sense that the perpendicular direction has a higher value. In order to conform to the experimental observation, the addition of a fairly strong anti-ferromagnetic mean field is required for this orientation. A value of $\lambda = -0.52$, the same value for all three compounds, produces the desired behavior of $\chi_1$ for this geometry as shown in fig. 6. Our experimental measurements of $\chi_1$ for CeCoIn$_5$ are also shown in this figure. Again, the agreement with the model is excellent. We emphasize that, while the mean-field parameter is altered on rotating the field, all other parameters for the three compounds remain equal to the values established from an analysis of the parallel geometry.

In conclusion, we have performed non-linear susceptibility measurements on CeCoIn$_5$ and have explained the results in a comprehensive manner with the single energy scale model of MM. Our approach has enabled an understanding of the magnetic properties of the related 115 heavy fermion systems CeRhIn$_5$ and CeIrIn$_5$. Although not explicitly demonstrated here, the same approach can be used to understand the behavior of these systems under pressure.

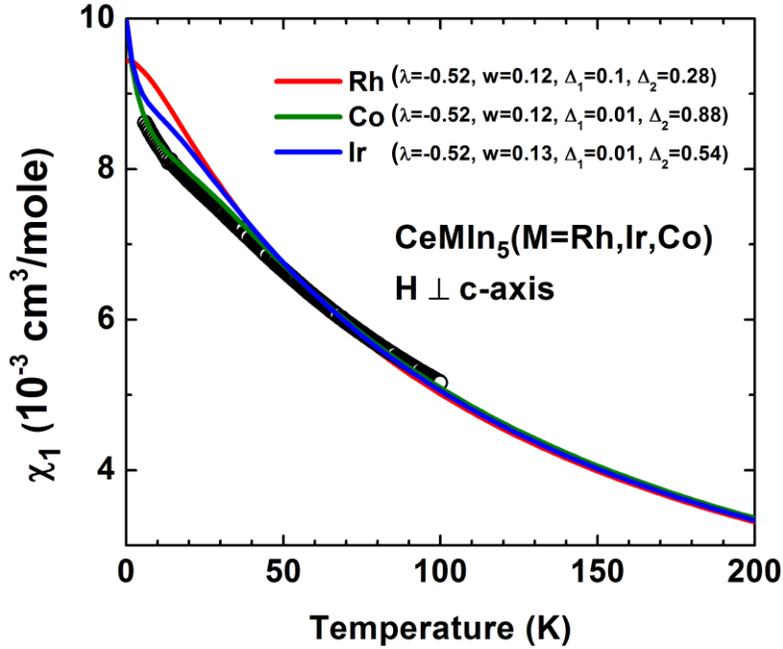

Figure 6: Shows the linear magnetic response for the perpendicular case. The green line is a fit with parameters as explained in the text to the CeCoIn$_5$ data ( open circles - sample#1).

It is significant that a simple Hamiltonian expressed in this work is able to account for the third order susceptibility of CeCoIn$_5$ and predictions are made for its MM field and the fifth order susceptibility. Since there are no conduction electrons in Hamiltonian (1) an objection to our model is that it describes an insulator and not a metal. However, we note that the presence of the conduction electrons are indeed considered in the model, albeit in an indirect manner, through the hybridization parameter, $w$. The inclusion of '$w$' and '$\lambda$' does influence both the temperaure and field-dependence of the magnetic response. Demonstrating that transport properties also can be derived from the same effective Hamiltonian is a challenge, and implementing it would be a major step forward in our understanding of heavy fermion metals.

Several recent studies have pointed out that CeCoIn$_5$ is situated close to a magnetic instability and an avoided quantum critical endpoint[17]. Thus, it is easy to drive CeCoIn$_5$ into a long range ordered magnetic state through a small pressure and/or by varying its chemical composition. This is consistent with our model where the parameter '$w \sim \Delta_1$' oscillates around zero. The comparable values obtained for all the parameters in Hamiltonian (1) could also suggest that the CeMIn$_5$ system is different from other heavy fermion metamagnets such as CeRu$_2$Si$_2$ where a change in the fermi surface, from large to small, is well established. The present work also demonstrates how an experimentally determined quantity such as the magnetic susceptibility, traditionally known to sense the bulk or total response, can indeed be used to decipher different contributions. Through non-linear susceptibility measurements, we are able to disentangle and confirm the contributions coming primarily from a metamagnetic part and a second paramagnetic spin. Such an approach, when generalized to include even higher order susceptibilities such as $\chi_5$, could be a powerful arsenal in understanding the physics of complex magnetic materials[18].

**Acknowledgements:** Research at UCSD (crystal growth) was supported by the U. S. Department of Energy, Office of Basic Energy Sciences, Division of Materials Sciences and Engineering under Award Grant no. DE-FG02-04-ER46105. The Tata Institute of Fundamental Research is supported by DAE, Govt of India. This research was conceived and organized by Bellave Shivaram (University of Virginia).

[9] "A Model for Strongly Correlated Electrons: Application to UPt$_3$ and CeRu$_2$Si$_2$", Bellave S. Shivaram, to be published, 2015.

[10] "Anomalous superconductivity and field-induced magnetism in CeCoIn$_5$", T. P. Murphy, Donavan Hall, E. C. Palm, S. W. Tozer, C. Petrovic, and Z. Fisk, R.G. Goodrich, P. G. Pagliuso, J. L. Sarrao, and J. D. Thompson, Phys. Rev. **B65**, 100514(R), 2002.
"Unconventional heavy-fermion superconductor CeCoIn$_5$: DC magnetization study at temperatures down to 50 mK", T. Tayama, A. Harita, T. Sakakibara, Y.Haga, H. Shishido, R. Settai, and Y. Onuki, Phys. Rev.B, **65**, 180504(R), (2002).

[11] "Evidence for a New Magnetic Field Scale in CeCoIn$_5$", I. Sheikin, H. Jin, R. Bel, K. Behnia, C. Proust, J. Flouquet, Y. Matsuda, D. Aoki, and Y. Onuki, Phys. Rev. Lett., **96**, 077207 (2006).

[12] "Magnetic and Thermal Properties of CeIrIn$_5$ and CeRhIn$_5$", Tetsuya Takeuchi, Tetsutaro Inoue, Kiyohiro Sugiyama, Dai Aoki, Yoshihumi Tokiwa, Yoshinori Haga, Koichi Kindo and Yoshichika Onuki, Journal of the Physical Society of Japan, **70,** 877–883, 2001. However, there is also a report of a MM transition at ~30 T in CeIrIn$_5$ and a ferromagnetic hysteresis in the 42 T range. See "Magnetic Transitions in CeIrIn$_5$", E.C. Palm, T.P.Murphy, Donovan Hall, S.W. Tozer, R.G. Goodrich and J.L. Sarrao, Physica, **B329-333**, 587,(2003).

[13] "Magnetism and Superconductivity in CeMIn$_5$ single crystals (M=Rh, Co)", S. Majumdar, G.Balakrishnan, M.R.Lees and D.K. McPaul, Acta Physica Polonca, 34, 1043, 2003.

[14] There is also an antiferromagnetic transition that immediately follows the maximum at a slightly lower temperature.

[15] In fig.1 we used the same scaling factor, T=85t, for all three compounds. Employing slightly different scaling factors for each of them maybe more appropriate but we ignore this refinement for the time being.

[16] We ignore the possibility of the weak metamagnetic signature seen at 50 T for the perpendicular geometry in CeRhIn$_5$. Such a feature is not present in our current model but can possibly be built in with further refinement to the model.

[17] "Avoided antiferromagnetic order and QCP in CeCoIn$_5$", A. Bianchi, R. Movshovich, I. Vekhter, P.G. Pagliuso, J.L. Sarrao, Phys. Rev. Lett., **95,** (2003), 257001.
"Reemergent Superconductivity and Avoided Quantum Criticality in Cd-Doped CeIrIn$_5$ under Pressure", Y. Chen, W. B. Jiang, C. Y. Guo, F. Ronning, E. D. Bauer, Tuson Park, H. Q. Yuan, Z. Fisk, J. D. Thompson, and Xin Lu, Phys. Rev. Lett., **114,** 146403 (2015)
"Itinerant Versus Localized Heavy-Electron Magnetism", Shintaro Hoshino and Yoshio Kuramoto, Phys. Rev. Lett., **111,** 026401 (2013).